\documentclass[aps,prd,twocolumn,superscriptaddress,amsfont,amssymb,amsmath,nofootinbib,showpacs,balancelastpage]{revtex4-1}

\usepackage{graphicx,longtable,natbib,mathrsfs,color}
\usepackage{txfonts,aas_macros}

\newcommand{\bs}{\boldsymbol}

\newcommand{\lb}{\left\langle}
\newcommand{\rb}{\right\rangle}

\begin{document}

\addtolength{\hoffset}{-0.525cm}
\addtolength{\textwidth}{1.05cm}
\title{Nonlinear ${\bs E}$-mode clustering in Lagrangian space}

\author{Hao-Ran~Yu}\email{haoran@cita.utoronto.ca}
\affiliation{Kavli Institute for Astronomy and Astrophysics, Peking University, Beijing 100871, China}
\affiliation{Canadian Institute for Theoretical Astrophysics, University of Toronto, 60 St. George Street, Toronto, Ontario M5S 3H8, Canada}

\author{Ue-Li~Pen}\email{pen@cita.utoronto.ca}
\affiliation{Canadian Institute for Theoretical Astrophysics, University of Toronto, 60 St. George Street, Toronto, Ontario M5S 3H8, Canada}
\affiliation{Dunlap Institute for Astronomy and Astrophysics, University of Toronto, 50 St. George Street, Toronto, Ontario M5S 3H4, Canada}
\affiliation{Canadian Institute for Advanced Research, CIFAR Program in Gravitation and Cosmology, Toronto, Ontario M5G 1Z8, Canada}
\affiliation{Perimeter Institute for Theoretical Physics, 31 Caroline Street North, Waterloo, Ontario, N2L 2Y5, Canada}

\author{Hong-Ming Zhu}\email{hmzhu@nao.cas.cn}
\affiliation{Key Laboratory for Computational Astrophysics, National Astronomical Observatories, Chinese Academy of Sciences, 20A Datun Road, Beijing 100012, China}
\affiliation{University of Chinese Academy of Sciences, Beijing 100049, China}

\date{\today}

\begin{abstract}
We study the nonlinear $E$-mode clustering in Lagrangian space
by using large scale structure $N$-body simulations
and use the displacement field information in Lagrangian space
to recover the primordial linear density field.
We find that, compared to Eulerian nonlinear density fields,
the $E$-mode displacement fields
in Lagrangian space improves the cross-correlation scale $k$ with
initial density field by a factor of 6-7,
containing two orders of magnitude more primordial information.
This illustrates ability of potential density reconstruction algorithms,
to improve the baryonic acoustic oscillation measurements
from current and future large scale structure surveys.

\end{abstract}


\maketitle

\section{Introduction}\label{sec.intro}
Our Universe starts from primordial Gaussian perturbations at a very early stage, 
and from those fluctuations, the gravitational instability drives the formation of 
the large scale structure (LSS) distribution of matter
\citep{1970A&A.....5...84Z,1985ApJ...292..371D}.
These structures grow 
linearly until the perturbations are large enough so that the first order 
perturbation theories are unable to analytically describe the LSS distributions 
\citep{2016JCAP...01..043M}. 
As a result, the final nonlinear LSS distribution contains higher order 
statistics, and thus makes it more challenging to be interpreted into basic
cosmological parameters. One such example is that, the baryonic acoustic oscillation (BAO)
scale can be used as a ``standard ruler'' to constrain the cosmic expansion history
and thus probes the dark energy properties \citep{2005NewAR..49..360E}, but
nonlinear evolution smears the BAO features and lowers the measurement
accuracy \citep{2005ApJ...633..560E,2012MNRAS.419.2949N}.
There are various attempts to recover earlier stages
of LSS, in which statistics are closer to Gaussian 
\citep{1992MNRAS.254..315W,2013MNRAS.436..759H}.
Because Gaussian fields can be adequately described by two-point statistics,
ideally after some recovery algorithms, more information can be extracted,
more straightforwardly, by power spectra or two-point correlation functions
\citep{2005MNRAS.360L..82R,2012MNRAS.421..832Y}.

Standard BAO reconstruction algorithms apply in Eulerian space.
They usually smooth the nonlinear density field on
linear scale ($\sim$10 Mpc$/h$) and reverse the large 
scale bulk flows by a negative Zel'dovich linear displacement
\citep{2007ApJ...664..675E,2009PhRvD..80l3501N,2009PhRvD..79f3523P}.
In the linear Lagrangian perturbation theory (LPT), the 
negative divergence of the displacement field
$-\nabla_q\cdot\bs\Psi(\bs q)$ with respect to Lagrangian coordinates $\bs q$
gives the linear density field \citep{2010PhDT.........4J},
and in many literatures
\cite{2013MNRAS.428..141N,2014PhRvD..89h3515C,2016JCAP...03..017B}
$\nabla_q\cdot\bs\Psi(\bs q)$ is compared with various standard LPT
and corrected LPT initial conditions. Their motivation was
trying to correct or modify the LPT displacement fields
in order that the final positions of particles are brought
closer to $N$-body results. Because of the absence
of Lagrangian space information in observations, few density
reconstruction algorithms are developed according to the relation
between displacement and linear density. However,
there are techniques to estimate the displacement field from a final 
distribution of matter.
For example, when a homogeneous initial matter distribution 
is assumed, there is a unique solution of curl-less displacement field to relate 
the initial and final distributions without shell crossing. This solution can be 
solved by a metric transformation equation
\citep{1995ApJS..100..269P,1998ApJS..115...19P}.
In the one-dimensional (1D) case, this solution 
simplifies to an ordering of mass elements by their final, Eulerian coordinates.
Zhu {\it et al.} \cite{2016arXiv160907041Z} apply this algorithm to the result
of a 1D simulation \citep{2016JCAP...01..043M} and obtain an estimated
displacement field $\tilde\Psi(q)$, and find
that this new method well recovers the linear
information and reconstructs 1D BAO peak in the correlation function.
In 3D and more realistic cases, one needs to
carefully consider effects of curl, shell crossing, 
complicated baryonic physics and biased tracers (galaxies).

\begin{figure}
 \centering
  \includegraphics[width=0.95\linewidth]{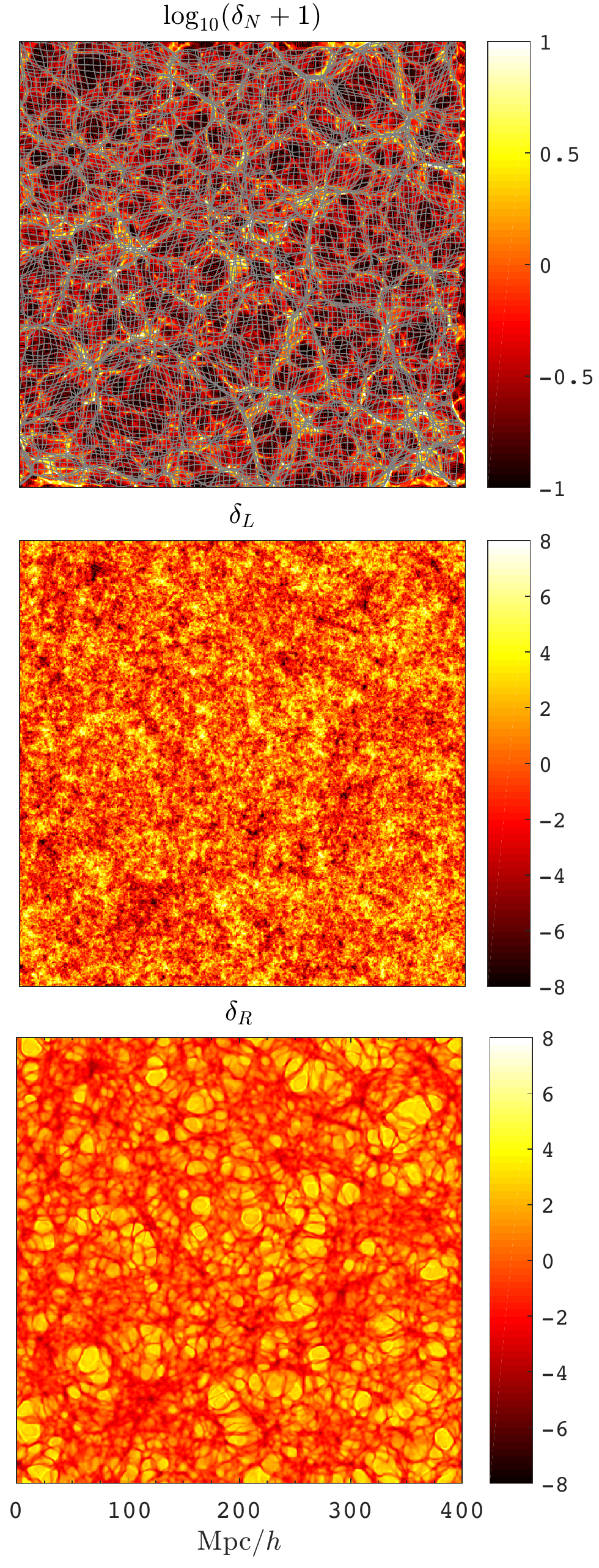}
  \caption{Visualization of the nonlinear density field $\delta_N$ (top),
  linear density field $\delta_L$ (middle) and the raw recovered density
  field $\delta_R$ (bottom). These projections have 9.375 Mpc$/h$ thickness
  and 400 Mpc$/h$ per side. The top panel shows the nonlinear displacement
  field $\bs\Psi$ by the deformed mesh, which traces the LSS of $\delta_N$.}
  \label{fig.projection}
\end{figure}

Before these steps, we need to quantify the amount of linear information
that can be recovered\footnote{To avoid ambiguity, we call
$\delta_R\equiv-\nabla_q\cdot\bs\Psi(\bs q)$ recovered linear density field, which
requires $\bs\Psi(\bs q)$ from $N$-body simulations, while
$-\nabla_q\cdot\tilde{\bs \Psi}(\bs q)$ is called the
reconstructed density field, which uses the estimated displacement field.}
from the full nonlinear displacement field $\bs\Psi(\bs q)$,
This field can be decomposed into a curl-less ``$E$-mode'' component
and a divergence less ``$B$-mode'' component, $\bs\Psi=\bs\Psi_E+\bs\Psi_B$,
where the $B$-mode is raised by late stage non-Gaussianities and is not
dominant in $\bs\Psi$ \citep{2014PhRvD..89h3515C}.
In $\bs\Psi(\bs q)$ from $N$-body simulations,
(1) $E$-mode displacement field is used for recovering the linear information
(although $B$-mode is also present); (2) shell crossing effects are fully considered;
(3) the setup is clean in absence of baryons and realistic
observables. Further reconstructions by $\tilde{\bs \Psi}(\bs q)$
can be compared with this result.
Furthermore, we study the scale dependent cross correlation
between $-\nabla_q\cdot\bs\Psi(\bs q)$ and linear density field
and construct optimal Wiener filters to get the optimal filtered
recovered linear density field and recovered linear power spectrum.

In the rest of the paper, we describe the simulation,
density recovery algorithm and Wiener filter setups in section
\ref{sec.method}. We show results in section \ref{sec.results}.
Discussion and conclusion are in section \ref{sec.discussion}.

\section{Method}\label{sec.method}
We show the LSS simulation and displacement field setups in section \ref{ss.sim}.
In section \ref{ss.reco}, we recover the linear density field from the
displacement field $\bs\Psi(\bs q)$ from simulations.
Note that potential reconstruction algorithms are
based on an estimated displacement field $\tilde{\bs \Psi}(\bs q)$
instead of $\bs\Psi(\bs q)$. In the following sections we
use $\delta_R$ to label the recovered linear density field
from $\bs\Psi(\bs q)$.

\subsection{Simulation}\label{ss.sim}
We use the open source cosmological simulation code {\tt CUBE} 
\citep{cafcube}.
Cosmological parameters are set as
$\Omega_m=0.27$, $\Omega_\Lambda=0.73$, $h_0=0.68$, $n_s=0.96$ and $\sigma_8=0.83$.
Initial conditions are generated at redshift $z=50$ 
using Zel'dovich approximation, and using a CLASS transfer function
\citep{2011JCAP...07..034B}.
$N_p=512^3$ $N$-body particles are evolved via 
their mutual gravitational interactions to $z=0$, in a periodic box with $L=400$ 
Mpc$/h$ per side. The code is set to use standard a particle-mesh (PM) algorithm 
\cite{1988csup.book.....H} on a two-level mesh grids
(for details see \cite{2013MNRAS.436..540H}) and cloud-in-cell
(CIC) is used in particle interpolations in force 
calculation and obtaining the density field $\rho({\bs x})$ in Eulerian coordinates 
${\bs x}$ at late stages. We use density contrast $\delta\equiv\rho/\lb\rho\rb-1$ 
to describe the density fluctuations. The primordial linear density field $
\delta_L$ is given by the initial stage and scaled to $z=0$ by the linear growth 
factor. In the top and middle panels of Fig.1 we show projections of the nonlinear density field
$\delta_N$ given by the simulation and the linear density field $\delta_L$.
$\delta_N$ is obtained by the particle distribution at redshift $z=0$, and
the particles are interpolated by using the cloud-in-cell (CIC) algorithm.
Because $\delta_N$ is highly nonlinear and follows an approximate
log-normal distribution, we plot $\log_{10}(\delta_N+1)$ instead, and show
the color scale $\log_{10}(\delta_N+1)\in[-1,1]$ (or $\rho_N/\langle\rho_N\rangle\in[0.1,10]$)
only, for a better visualization.
The nonlinear evolution of $\delta_N$ makes it very different from $\delta_L$
in appearance.

\begin{figure}[t] \centering
  \includegraphics[width=1.0\linewidth]{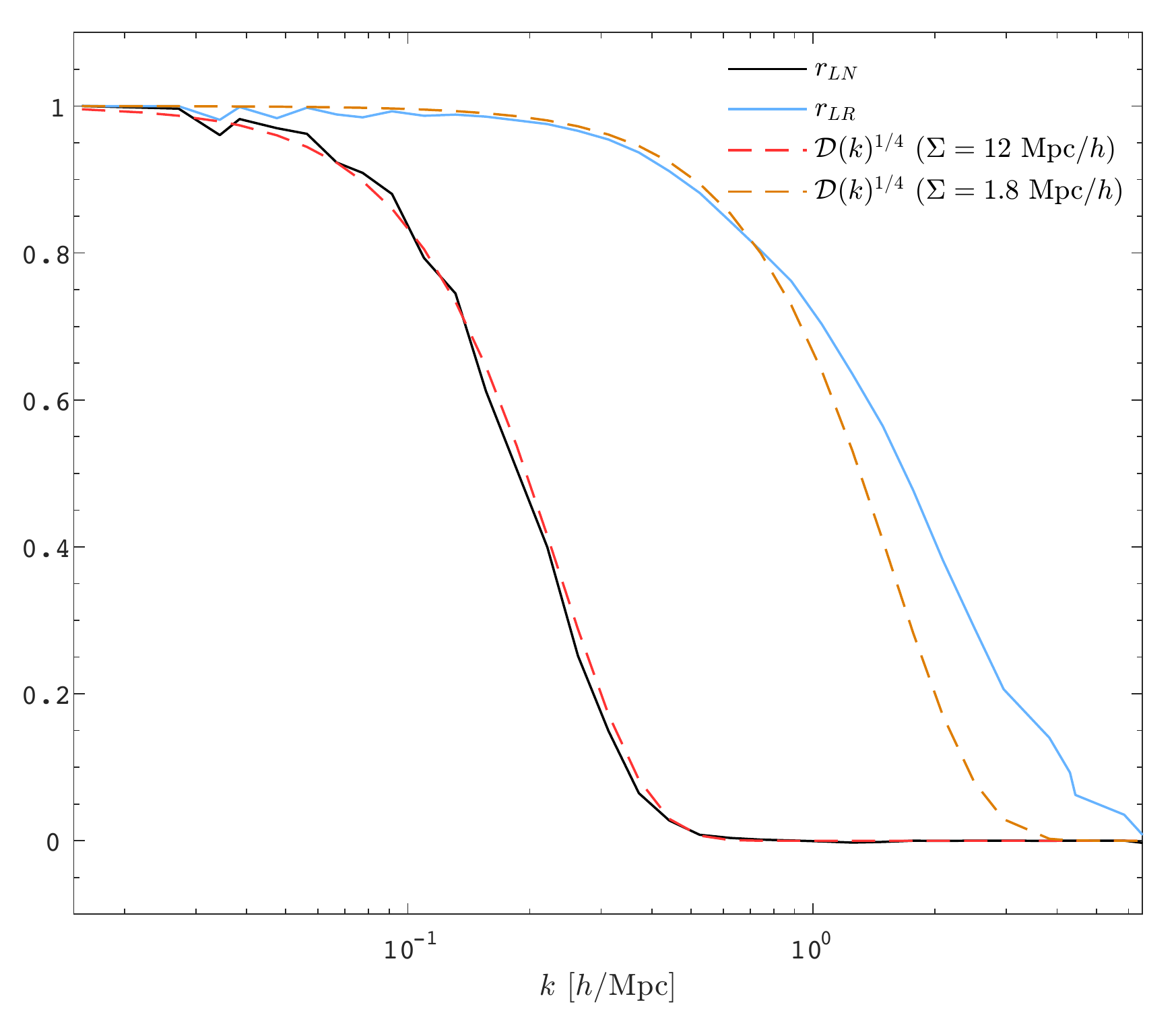}
  \caption{Correlation functions $r(\delta_L,\delta_N)$ and $r(\delta_L,\delta_R)$
  (solid lines) and their scaled BAO damping models (dotted lines).}
  \label{fig.corr}
\end{figure}

The two-point statistics of these density fields are quantified by the cross power 
spectrum $P_{ij}(k)\equiv(2\pi)^{-3}\langle|\delta_i(k)||\delta_j(k)|\rangle$, 
where subscripts $i,j$ may refer to linear ($L$), nonlinear ($N$),
recovered ($R$), or noise ($n$) density 
fields. When $i=j$ it reduces to the auto power spectrum $P_{ii}(k)$ or $P(k)$. We 
usually plot the dimensionless power spectrum $\Delta^2(k)\equiv k^3P(k)/2\pi^2$. 

\subsection{Density recovery}\label{ss.reco}
In the simulation, we use particle-ID (PID) to record the initial (Lagrangian) location ${\bs 
q}$ of particles, and the information is tracked until the $z=0$ and we can get the 
Lagrangian displacement vector ${\bs \Psi}\equiv{\bs x}-{\bs q}$ for every 
particle. Then these vectors are interpolated onto the initial Lagrangian 
coordinates ${\bs q}$ of particles and we get the displacement field ${\bs \Psi}
({\bs q})$.
To visualize the $\bs\Psi$ field, we draw a 3D uniform mesh over the volume,
and use the given $\bs\Psi$ field to deform the mesh according to the direction
and physical amplitude of $\bs\Psi$. In the top panel of Fig.\ref{fig.projection},
The resulting mesh illustrates a ``pseudocurvilinear coordinate'' similar to \cite{1995ApJS..100..269P},
however the mesh can be overlapped due to shell crossing. The densest mesh grids
trace the densest structures of $\delta_N$, whereas the undeformed grid positions
are the Lagrangian coordinates in which we do the density recovery.
The raw recovered density field is given by the differential motion of matter 
elements,
\begin{equation}
    \delta_R=-\nabla\cdot{\bs \Psi}({\bs q}).
\end{equation}
Because the recovery processes are implemented on Lagrangian coordinates,
$\delta_R$ takes the coordinates of $\bs q$ instead of $\bs x$.
We just write $\bs q$'s Fourier wave number
$k_q$ as $k$ to simplify the expression.

To quantify the linear information in the recovered density field $\delta_R$,
we decompose $\delta_L$ in Fourier space into two uncorrelated parts,
\begin{equation}\label{eq.decompose}
    \delta_L(k)=r'\delta_R+\delta_n,
\end{equation}
where the first term $r'\delta_R$ is completely correlated with $\delta_R$,
meaning the linear information we can extract from $\delta_R$.
The noise term $\delta_n$ is uncorrelated with $\delta_R$, being
the rest of linear information which is not contained in $\delta_R$.
Correlating equation (\ref{eq.decompose}) with $\delta_R$ gives
\begin{equation}
    P_{LR}=r'P_{RR}+P_{nR},
\end{equation}
where $P_{ij}\equiv\langle\delta_i\delta_j\rangle$ denotes the cross power 
spectrum. Since $\delta_n$ is uncorrelated with $\delta_R$, $P_{nR}=0$. With the 
definition of cross correlation coefficient $r(\delta_L,\delta_R)\equiv P_{LR}/\sqrt{P_{LL}P_{RR}}$ 
and bias $b^2=P_{RR}/P_{LL}$, we solve $r'=P_{LR}/P_{RR}=rb^{-1}$.
Note that these computations above and below can also be applied on $\delta_N$ for
comparison, by replacing ``$_R$'' with ``$_N$'' in the equations,
while we do not rewrite them explicitly
in the paper for simplicity. From these, we plot the cross 
correlation coefficient $r_{LN}=r(\delta_L,\delta_N)$ and
$r_{LR}=r(\delta_L,\delta_R)$ in Fig.\ref{fig.corr}.
$r_{LN}$ shows no correlation starting from $k\simeq 0.4 h/$Mpc \citep{2004MNRAS.355..129S}.
Clearly, $
\delta_R$ contains much more linear information on smaller scales.

According to equation (\ref{eq.decompose}), the auto power spectrum of $\delta_L$ is decomposed as
\begin{equation}\label{eq.power}
    P_{LL}=r^2b^{-2}P_{RR}+P_{nn},
\end{equation}
and $P_{nn}=(1-r^2)P_{LL}$. We also explicitly
compute the cross power spectrum between $\delta_R$ and $\delta_n=\delta_L-rb^{-1}\delta_R$,
and found that $rb^{-1}P_{nR}$ is about two orders of magnitude lower than $P_{LL}$,
being consistent with zero. This confirms that
the signal term $rb^{-1}\delta_R$ and the noise term $\delta_n$
is indeed uncorrelated and validates equation (\ref{eq.power}).
According to these two terms, we construct a Wiener filter to filter
out the uncorrelated part in $\delta_R$:
\begin{equation}
    W(k)=\frac{r^2b^{-2}P_{RR}}{r^2b^{-2}P_{RR}+P_{nn}}=r^2.
\end{equation}
The optimal recovered density is given by
\begin{equation}
    \tilde\delta_R=Wb^{-1}\delta_R,
\end{equation}
and the optimal recovered power spectrum is given by
\begin{equation}\label{eq.opt}
    \tilde P=W^2b^{-2}P_{RR}=W^2P_{LL}+W^2b^{-2}P_{NN}.
\end{equation}
Here $W^2$ describes the damping of the linear power spectrum.

\section{Results}\label{sec.results}

To visualize the above algorithms, a projection of $\delta_R$ is plotted in
the bottom panel of Fig.\ref{fig.projection}, which looks closer to 
$\delta_L$ compared to $\delta_N$. However the smallest scale structures are unable
to be recovered.

As discussed in section \ref{ss.reco},
Fig.\ref{fig.corr} shows the cross correlation functions
$r_{LN}$ and $r_{LR}$. The latter extends the correlation
with $\delta_L$ to smaller scales by nearly an order of
magnitude. The extra correlation scales well cover the BAO
scales of our interest.

\begin{figure}[t] \centering
  \includegraphics[width=1.0\linewidth]{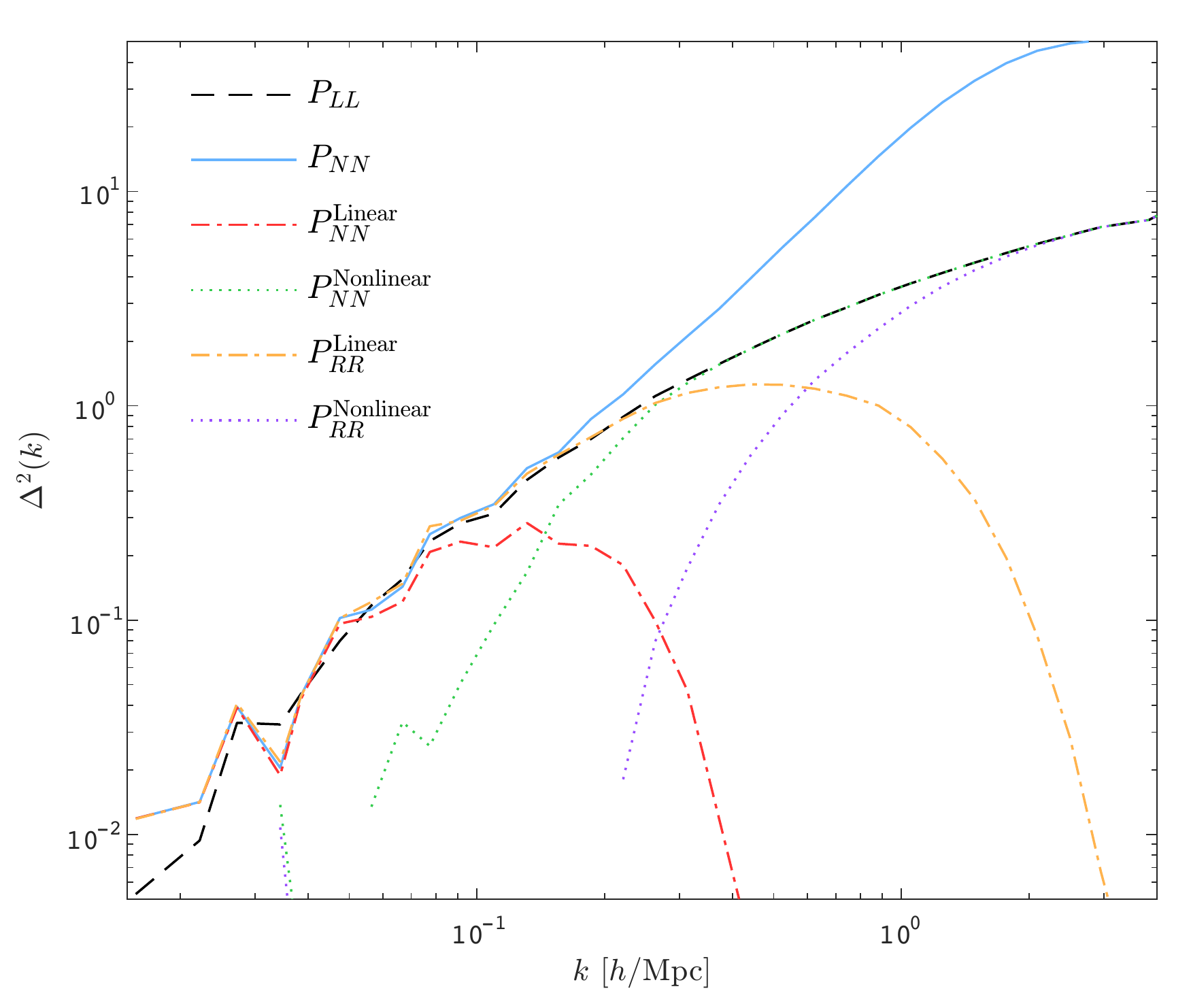}
  \caption{Power spectra of $\delta_L$, $\delta_N$, and the decomposition
  of $P_{LL}$ into correlated parts and noise terms according to equation
  (\ref{eq.power}).}
  \label{fig.recopower}
\end{figure}

\begin{figure}[t] \centering
  \includegraphics[width=1.0\linewidth]{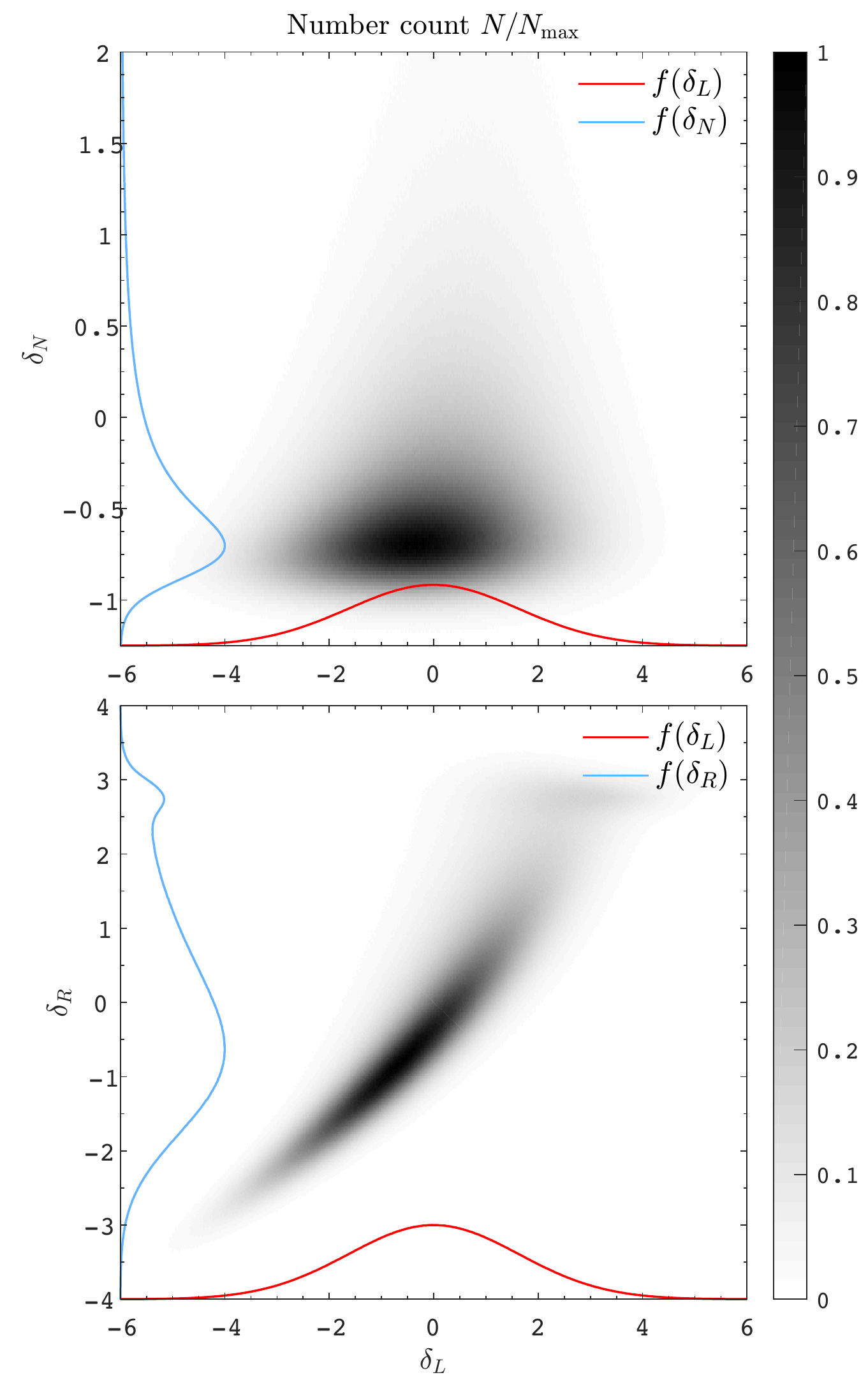}
  \caption{Probability distribution functions (PDFs)
  $f(\delta_L,\delta_N)$ and $f(\delta_L,\delta_R)$, showing in the upper
  and lower panels respectively. Both red curves on the $x$-axes show $f(\delta_L)$, following
  a Gaussian distribution. The two curves on the $y$-axes in the upper and the lower panel
  show respectively $f(\delta_N)$ and $f(\delta_R)$.}
  \label{fig.pdfs}
\end{figure}

In Fig.\ref{fig.recopower}, we show the auto power spectra of
$\delta_L$ and $\delta_N$ in black dashed and blue solid curves.
Their difference shows the nonlinear evolution of LSS on 
small scales. Their cross power (not shown for clarity of the figure)
drops to a very low value on scales $k\gtrsim 0.1h/$Mpc,
indicating a loss of linear information in the
nonlinear power spectrum $P_{NN}$. This scenario directly
leads to how $P_{LL}$ is decomposed according to equation
(\ref{eq.power}). In nonlinear case (with ``$_R$'' replaced
by ``$_N$'' equation (\ref{eq.power})), on small
scales $k\gtrsim 0.1h/$Mpc, $P_{LL}$ is dominated by
uncorrelated, nonlinear noise, shown in the green
dotted line. In the case of $\delta_R$, however, $P_{LL}$
is decomposed into the orange dash-dotted correlated
part and the purple dotted uncorrelated part according to
equation (\ref{eq.power}). The correlated power spectrum
is dominated on BAO scales of our interest.

To quantify the improvement of cross-correlation in the power
spectrum, we compute the 
damping factors $W^2(k)$ respectively for the optimal filtered 
nonlinear and recovered density fields $\tilde\delta_N$
and $\tilde\delta_R$.
We fit Gaussian BAO damping models
${\mathcal D}(k)=\exp(-k^2\Sigma^2/2)$ to these $W^2(k)$'s
and give $\Sigma=1.8$ Mpc$/h$ and $\Sigma=12$ 
Mpc$/h$ for $\delta_R$ and $\delta_N$.
Since ${\mathcal D}(k)=W^2=r^4$, we plot
${\mathcal D}_N^{1/4}$ and ${\mathcal D}_R^{1/4}$
over $r_{LN}$ and $r_{LR}$ in Fig.\ref{fig.corr}.
The analyses are repeated with various box sizes (100, 300, 800 Mpc$/h$ per side)
and give consistent results.

To further illustrate the improvement in real space
one point function correlations,
in Fig.\ref{fig.pdfs} we use the probability distribution
as functions (PDFs) of $(\delta_L,\delta_N)$ and $(\delta_L,\delta_R)$
to show the point-point correlation between these two pairs of density
fields. Since $\delta_n$ in equation (\ref{eq.decompose}) is
uncorrelated, we use Wiener filtered fields. To keep the
consistency over $\delta_L$, $\delta_N$ and $\delta_R$, we
use the $W(k)=r^2_{LR}$ as the Wiener filter.
The grey-scaled plots in the center of both panels show the two-variable PDFs,
whereas their projections onto each variable are just
one-variable PDFs -- $f(\delta_L)$, $f(\delta_N)$ and $f(\delta_R)$, shown
as red/blue curves on the axes of Fig.\ref{fig.pdfs}.
In the top panel, $\delta_N$ shows an approximate log-normal
distribution (blue curve) and $\delta_L$ follows an expected
Gaussian distribution. They show tiny positive correlation in the
2D PDF.
Because in Fourier space, $\delta_L$ and $\delta_N$ have
correlations on only very large, linear scales (Fig.\ref{fig.corr}),
they result in little correlation in real space -- initial density fluctuations
in Lagrangian coordinates are evolved/transformed to Eulerian coordinates.
As the recovery is done in Lagrangian space, it recovers
certain amount of correlation, as shown in the 2D PDF of the bottom panel
of Fig.\ref{fig.pdfs}. One can also see that $\delta_R$ follows
a much closer Gaussian distribution (blue curve of the bottom panel).
In denser regions of $\delta_L$, $\delta_R$ is saturated at
$\delta_R=3$, signifying the extreme collapsing of matter \citep{2013MNRAS.428..141N}:
$\delta_R=-\nabla\cdot\bs\Psi=\nabla\cdot\bs q=3$. Shell crossing
makes $\delta_R$ oscillate around 3. These second uncorrelated peaks
damp out as we go to higher redshifts.

\section{Discussion and conclusion}\label{sec.discussion}
We extract the actual displacement field of matter elements in cosmological $N$-body
simulations, and use this displacement field to study the LSS nonlinear
clustering in Lagrangian space. The displacement information is used to recover
the primordial linear perturbations. The result shows a prominent improvement from
$r_{LN}$ to $r_{LR}$ in Fig.\ref{fig.corr} -- recovering the lost linear information on
nearly an order of magnitude smaller scales.
This is achieved by implementing differential movement information
of matter elements on Lagrangian coordinates, rather than on
Eulerian coordinates. This result illustrates the feasibility
of using estimated displacement field $\tilde{\bs \Psi}(\bs q)$
to reconstruct primordial linear density field.
A straightforward example of a estimation of $\tilde{\bs \Psi}(\bs q)$
is given by \cite{1995ApJS..100..269P,1998ApJS..115...19P}.
In reality, one needs
to consider all aspects including vorticity, shell crossing, bias, noise
and data complexities. The impact of these factors can be quantitatively
compared with the impact of different estimation methods, and with
the exact solution by $N$-body simulations.

The advantage of using displacement field in reconstruction is
its insensitive response from highly nonlinearities.
Nonlinear densities $\delta_N$ can be arbitrarily large -- 
one expects virialized regions to be observable, where nonlinear
density is given by the inverse determinant of the Jacobian
$\delta_N=|\bs J|^{-1}\gtrsim 200$.
However, reconstructed densities $\delta_R$ are given by
the trace $\rm{tr}(\bs J)$ and saturates at 3.
Actually, the displacement fields are dominated by early stage linear processes,
which is the Lagrangian-Eulerian coordinate transform,
while late stage shell crossing, nonlinear and baryonic processes
only fine-tune the final position $\bs x$
\citep{2014PhRvD..89h3515C}. Compared with estimated displacement fields $\tilde{\bs \Psi}(\bs q)$
by \cite{1995ApJS..100..269P}, which do not have shell crossing,
the additional shell crossing information in $\bs\Psi(\bs q)$
is uncorrelated with $\delta_L$. This insensitive response from
nonlinearities enables the stability of reconstruction algorithms,
which are expected to give similar results of this paper. 
In contrast, traditional
treatments in reconstruction deals directly on density fields which sensitively
relies on nonlinear processes --  density values can vary by orders
of magnitude due to nonlinear/baryonic physics and many sources of errors.

\section*{ACKNOWLEDGMENTS}
We thank Pengjie Zhang and Kwan Chuen Chan for helpful discussions and comments.
We acknowledge funding from NSERC.
H.R.Y. acknowledges General Financial Grant No. 2015M570884 and Special Financial Grant No. 2016T90009 from the China PostdoctRoal Science Foundation.
H.M.Z. acknowledges the support of the Chinese MoST 863 program under Grant
No. 2012AA121701, the CAS Science Strategic Priority Research Program
XDB09000000 and the NSFC under Grant No. 11373030.
Research at the Perimeter Institute is supported by the Government of Canada
through Industry Canada and by the Province of Ontario through the Ministry of
Research $\&$ Innovation.

\bibliographystyle{h-physrev3}
\bibliography{../haoran_ref}

\end{document}